\documentclass{hotnets22}

\usepackage{times}  
\usepackage{hyperref}
\usepackage{titlesec}
\usepackage{listings}

\hypersetup{pdfstartview=FitH,pdfpagelayout=SinglePage}

\newcommand{\tool}{\textsc{CoSynth}}

\begin{document}

\conferenceinfo{Submitted to HotNets 2023} {}


\title{What do LLMs need to Synthesize Correct Router Configurations?}
 \author{Rajdeep Mondal \\ \href{mailto:mondalrajdeep14@ucla.edu}{mondalrajdeep14@ucla.edu}
     \and Alan Tang \\ \href{mailto:atang42@cs.ucla.edu}{atang42@cs.ucla.edu}
     \and Ryan Beckett \\ \href{mailto:Ryan.Beckett@microsoft.com}{Ryan.Beckett@microsoft.com}
     \and Todd Millstein \\ \href{mailto:todd@cs.ucla.edu}{todd@cs.ucla.edu}
      \and George Varghese \\ \href{mailto:varghese@cs.ucla.edu}{varghese@cs.ucla.edu}}

\maketitle

\begin{abstract}

We investigate whether Large Language Models (e.g., GPT-4) can synthesize correct router configurations with reduced manual effort. We find GPT-4 works very badly by itself, producing promising draft configurations but with egregious errors in topology, syntax, and semantics.  Our strategy, that we call {\em Verified Prompt Programming}, is to combine GPT-4 with verifiers, and use localized feedback from the verifier to automatically correct errors.
Verification requires a specification and actionable localized feedback to be effective.  We show results for two use cases:  translating from Cisco to Juniper configurations on a single router, and implementing no-transit policy on multiple routers. 
While human input is still required, if we define the {\em leverage} as the number of automated prompts to the number of human prompts, our experiments show a leverage of 10X for Juniper translation, and 6X for implementing no-transit policy, ending with
verified configurations.

\end{abstract}

\section{Introduction}

\begin{quote}
The limitations of his knowledge were as startling as its profundity -- {\em Hardy on Ramanujam}~\cite{hardy}
\end{quote}

While GPT-4 and other language models have shown great success in some domains (e.g., writing poems, passing the LSAT) they have been shown to have issues in other domains (e.g,. math, word puzzles)~\cite{sparksofagi}. Language models have had some success in helping users write sequential programs in systems like CoPilot~\cite{copilot} and Jigsaw~\cite{jigsaw}.  Our paper examines GPT-4's ability to write router configuration files, traditionally written by humans, that help tune routes and forwarding decisions and are critical for network operation. Our early experiments show that GPT-4 by itself is an
``idiot-savant", capable of brilliance but also making simple errors that an operator would be fired for making.

Critics have derided large language models (LLMs) as mere ``stochastic parrots''~\cite{gebru},because they produce text (say of a program) {\em syntactically} by predicting the next word based on a statistical model derived by training on a vast corpus of text found on the Internet. Our broader goal beyond synthesizing configs is to see whether LLMs can be combined with other {\em programs} (via APIs) to come closer to a ``stochastic owl'' that understands program semantics.

A plausible way to introduce semantics is to pair a LLM with an automatic verifier such as a SAT solver or a model checker. But verification is not a panacea.  First, a  verifier cannot prove correctness without a specification.  In practice,
specifications are incomplete, so not all solutions are in fact acceptable to the user. 
Second, for the verifier to automatically (with minimal human aid) interact with the LLM, the verifier must provide actionable feedback.  We found it was easier for the LLM to correct itself using feedback from {\em modular verification} of components of a network (individual routers~\cite{lightyear} or even route maps within a router~\cite{campion}), rather than the network as a whole.


\begin{figure}[htb]
    \centering
    \includegraphics[width=\linewidth]{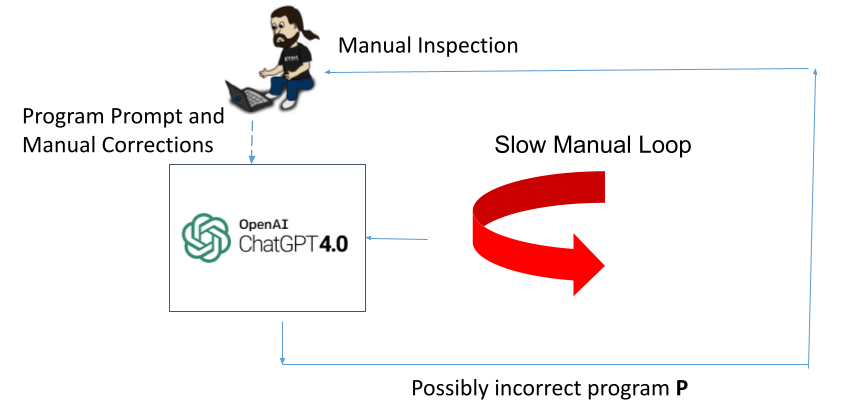}
    \vspace{-3em}
    \caption{Pair Programming using human correction.}
    \label{fig:pairprogramming}
\end{figure}

Figure~\ref{fig:pairprogramming} shows the traditional method of {\em pair programming (PP)}, embodied in systems like GitHub CoPilot~\cite{copilot}, where a human and an AI work together to author a program.  In pair programming, the AI and the human form a tuple $(A, H)$ and the human $H$ {\em manually} checks for correctness of the output of the AI $A$ and then {\em manually} issues correction prompts to $A$ as shown in the figure.  Such manual initial prompting and subsequent manual correction is often called {\em prompt engineering}.

\begin{figure}[htb]
    \centering
    \includegraphics[width=\linewidth]{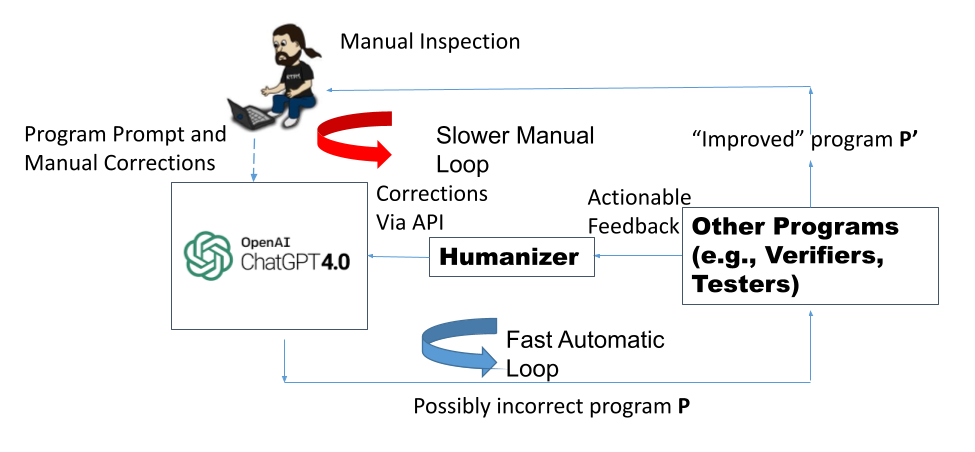}
    \vspace{-2.8em}
    \caption{Verified prompt programming.}
    \label{fig:verifiedpairprogramming}
\end{figure}

Figure~\ref{fig:verifiedpairprogramming} shows our alternate vision. In what we call {\em Verified Prompt Programming (VPP)}, the AI, the human, and a verification suite ($V$)  form a triple $(A, H, V)$. The verification suite checks for correctness and automatically issues localized corrections.
$V$ may abandon automatic correction after some number of trials, and the human must still correct manually. However, our hypothesis is that human effort is reduced as the output grows “closer” to a correct program.  

Notice that there is a fast inner loop between $V$ and $A$, where verifier results are automatically fed back to GPT-4.  Since verifier feedback is often cryptic, we use simple code that we call a {\em humanizer} that converts the feedback to natural language prompts that are given to GPT-4.  When $V$ either determines the final configuration is correct or a time bound elapses, $V$ sends the output back to the user as part of the slow manual loop.  We examine a ``reduced work hypothesis": that the work in the manual loop in  Figure~\ref{fig:verifiedpairprogramming} is significantly less than then the manual work in  Figure~\ref{fig:pairprogramming}

To quantify reduced human effort we introduce a simple measure that may be useful in other VPP contexts.  
Define {\em leverage} as the ratio $L$ of the number of automated prompts in Figure~\ref{fig:verifiedpairprogramming} to the number of human prompts. Leverage measures the effect of the {\em verifier suite}, the 
potential improvement in going from $(A, H)$ to $(A, V, H)$, keeping the language model $A$ and the human $H$ the same. 

The reader may think the real leverage is the improvement from $H$ to $(A, H)$, or from $H$ to $(A, V, H)$. But this depends on the capability of the human $H$ and is hard to make uniform or repeatable.  Given how error-prone
$(A, H)$ is for configurations, we find it more natural to measure the improvement caused by VPP. If a future LLM, say GPT-6, produces near-perfect configurations, leverage will decrease as there is less need
for automatic correction.  Our definition also assumes every automatic correction in Figure~\ref{fig:verifiedpairprogramming}
would otherwise be be done by a human in Figure~\ref{fig:pairprogramming}. 

The reduced work hypothesis is that the leverage $L > 1$ is high.  Even if the leverage is low (say $1$), since it is crucial that router configurations be correct, combining with a verifier seems critical.  We were happy to find that in both use cases we did end with verified configurations via GPT-4: this was not obvious at the outset.

This vision and hypothesis extends beyond synthesizing configs to more general programs. Prompt programming (as opposed to prompt engineering) also reflects the use of APIs and automatically generated feedback prompts that may be more generally useful.  However, network configs are a simple enough domain to experiment with. Further, there exist config verifiers (e.g., Campion~\cite{campion} and Lightyear~\cite{lightyear}) that provide actionable localized feedback.

For the rest of this paper, we examine the reduced manual work hypothesis and measure leverage for two use cases: translating a config on a single router from Cisco to Juniper syntax, and implementing a simple policy (``no transit") on a network of 7 routers. Section~\ref{sec:overview} describes the system organization of a potential system we call \tool.  Section~\ref{sec:ciscotojuniper} describes experiments with Cisco to Juniper translation, while Section~\ref{sec:localsynthesis} describes implementing no-transit on multiple routers.  Section~\ref{sec:previouswork} compares our ideas to previous work and Section~\ref{sec:conclusions} describes lessons learned.

\section{System Organization}
\label{sec:overview}

\begin{figure}[htb]
    \centering
    \includegraphics[width=\linewidth]{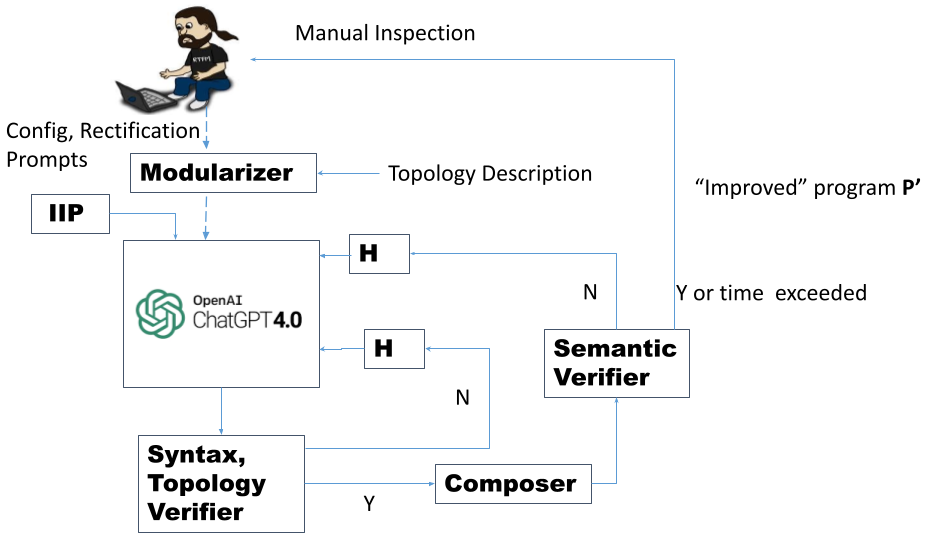}
    \vspace{-2em}
    \caption{Verified prompt programming for Configs}
    \label{fig:CoSynth3}
\end{figure}

Figure~\ref{fig:CoSynth3} is a refinement of the more general Verified Pair Programming (VPP) vision of Figure~\ref{fig:verifiedpairprogramming} that we call \tool. We emphasize we have not built \tool.  While we use GPT-4 we have not been able  to access the APIs, and so manually simulated the API calls with prompts to ChatGPT.  Our goal is not to demonstrate a working system but instead to explore GPT-4's ability to author configurations, as in the ``Sparks of AGI" paper~\cite{sparksofagi}.

The verification suite shown in  Figure~\ref{fig:CoSynth3} consists minimally of two verifiers, a syntax verifier (we used Batfish~\cite{batfish}) and a semantics verifier (we used different ones depending on the use case). For our second use case, we used a third verifier, a topology verifier (that we wrote in Python) as we found that GPT-4 sometimes missed announcing routes to neighbors.  The user provides a precise natural language description of the context (topology, routers, interfaces) and the desired task (e.g. the Cisco config and a request to translate it to Juniper). GPT-4 output is 
fed first to Batfish to check for syntax errors. \tool\ sends GPT-4 feedback about erroneous lines, ``humanized" in natural language (see Table~\ref{tab:translation-prompts} for examples). The boxes labelled {\bf H} in Figure~\ref{fig:CoSynth3} correspond to the humanizer in  Figure~\ref{fig:verifiedpairprogramming}.  

If all syntax errors are corrected (if too many syntax correction attempts occur, \tool\ punts to the user),  the output is passed to the semantics verifier.  For our first use case, we use Campion~\cite{campion} as a verifier. For our second use case we use Batfish's symbolic route map analysis as the verifier, asking it to verify local policies that together ensure the desired global policy, as in Lightyear~\cite{lightyear}.  Once again, the semantic verifier feedback is passed back, suitably humanized, to GPT-4.  We found that GPT-4 would sometimes correct a semantic error while introducing a new syntax error, in which case we had to return to the syntax verifier.  When the semantic verifier attests to a correct config or too many correction attempts transpire, \tool\ returns to the human.

When \tool\ works with multiple routers, we used another module called a ``Modularizer"  (Figure~\ref{fig:CoSynth3}).  For network configs, the idea is that we start with a precise machine readable (we use JSON) description of the ``modules" which in our case is the topology and the connections.  The Modularizer outputs a sequence of Natural Language Prompts that describes the topology to GPT-4 (e.g.,. Router $R1$ is connected to Router $R2$ via interface $I1$ at $R1$ and $I2$ at $R2$).  The Modularizer can also take a general specification of local policies (e.g. edge routers add a specific community on ingress) and output a specific local specification for each router for the semantic verifier. The Composer puts back the
pieces (in our case in a folder for Batfish).

The modularizer follows the prompt engineering paradigm "Give the Model Time to
Think"~\cite{chatgptlessons}, which suggests breaking a complex prompt into simpler sub-prompts. Exploiting modularity is a way to do so for program synthesis.  A second technique we find useful is what 
is called single shot prompting~\cite{chatgptlessons}. We start each chat with a set of  {\em initial instruction 
prompts (IIP)} (Figure~\ref{fig:CoSynth3}) loaded from a database for avoiding common mistakes.  The IIP database can be built and added by experts over time.  The I/O examples in Jigsaw~\cite{jigsaw} are an IIP, but our IIP contains instructions not examples.

\section{Cisco to Juniper Translation}
\label{sec:ciscotojuniper}

We translate a Cisco configuration into an equivalent Juniper one using Verified Pair Programming. Batfish~\cite{batfish} is used to identify syntax errors. Campion~\cite{campion} is used to detect and localize semantic differences that are used to refine the result. We show examples of the issues encountered, and discuss success and limitations of the approach.

\subsection{Method}

First, we provide the Cisco configuration, and the prompt: "Translate the configuration into an equivalent Juniper configuration." GPT-4 will produce a translation into Junos format that typically contains several errors and differences. We try to rectify these errors iteratively. First, we use Batfish~\cite{batfish} and Campion~\cite{campion}to detect any errors or differences, and then use a simple script to produce a prompt for GPT-4 (the humanizer $H$ in Figure~\ref{fig:CoSynth3}) to try and fix this error. After GPT-4 attempts to resolve the issue, we ask it to print the entire configuration and check the result using verification tools again. For our experiment we focus exclusively on behavior related to routing and forwarding, ignoring potentially important features such as NTP servers.

To design the humanizer (i.e., automatically generate a prompt informing GPT-4 of the errors present), we distinguish four classes of configuration errors:

\textbf{Syntax errors:} Batfish produces parse warnings identifying relevant lines that do not use valid Juniper syntax.

\textbf{Structural mismatch:} This is when a component, connection, or named policy is present in the original configuration but not in the translation (or is present in the translation but not the original). For example, if the original configuration defined a BGP neighbor but there is no corresponding neighbor in the translation, there would be a mismatch in the routing connections. Similarly, if there are corresponding BGP neighbor definitions in both configurations, but one configuration has an import policy defined while the other does not, that would be a mismatch in the named policies present. Campion is able to detect this, and identify the missing or extra items.

\textbf{Attribute differences:} This is when a numerical attribute has a different value between the two configurations. An example is OSPF link cost difference between two corresponding interfaces. Campion detects these and prints the attributes for corresponding components. 

\textbf{Policy behavior differences:} This is when a route map or access control list has a semantic difference. Route maps are used to filter incoming or outgoing route advertisements, so a difference would mean that that there are some route advertisements that are allowed by one router but not allowed by the other. Campion is able to detect these and output the relevant policy names, prefixes, and lines for these differences.

The distinction among errors helps for two reasons. First, syntax errors and structural mismatches have to be handled earlier since they can mask attribute differences and policy behavior differences. Second, different types of errors require different humanized prompts, while errors of the same type can reuse similar prompts. Each type of error can be summarized with a formulaic prompt with some fields inserted based on the error reported by Batfish or Campion. 

Table~\ref{tab:translation-prompts} shows the formulas and examples of generated prompts. Batfish parse errors and warnings can be reused as prompts for syntax errors. Prompts for structural mismatches and attribute differences are easily generated from the relevant components and attributes. Policy behavior differences are more difficult since it is not always clear how to describe the affected input space that is treated differently. We opt for the approach of giving an example prefix.

\begin{table}[tp]
\small
\begin{tabular}{|l|l|l}
\cline{1-2}
\textbf{Type} &
  \textbf{Generated Prompt} &
   \\ \cline{1-2}
\begin{tabular}[c]{@{}l@{}}Syntax\\ error\end{tabular} &
  \begin{tabular}[c]{@{}l@{}}There is a syntax error: \\ `\textit{policy-options prefix-list our-networks 1.2.3.0/24-32}'\end{tabular} &
   \\ \cline{1-2}
\begin{tabular}[c]{@{}l@{}}Structural\\ mismatch\end{tabular} &
  \begin{tabular}[c]{@{}l@{}}In the original configuration,\\ there is \textit{an import route map for bgp neighbor 2.3.4.5},\\ but in the translation,\\ there is no corresponding \textit{route map}\end{tabular} &
   \\ \cline{1-2}
\begin{tabular}[c]{@{}l@{}}Attribute\\ difference\end{tabular} &
  \begin{tabular}[c]{@{}l@{}}In the original configuration,\\ \textit{the OSPF link for Loopback0} has \textit{cost} set to \textit{1},\\ but in the translation, the corresponding \\ 
  \textit{link to lo0.0} has \textit{cost} set to \textit{0}\end{tabular} &
   \\ \cline{1-2}
\begin{tabular}[c]{@{}l@{}}Policy\\ behavior\\ difference\end{tabular} &
  \begin{tabular}[c]{@{}l@{}}In the original configuration, for \textit{the prefix 1.2.3.0/25}, \\ the \textit{BGP export policy to\_provider for BGP neighbor 2.3.4.5}\\ performs the following action: \textit{ACCEPT}.\\ But, in the translation,\\ the corresponding \textit{BGP export policy to\_provider}\\  performs the following action: \textit{REJECT}\end{tabular} &
   \\ \cline{1-2}
\end{tabular}
\vspace{-1em}
\caption{Sample rectification prompts for translation generated using formulas (non-italicized text), and fields generated from Batfish and Campion (italicized text).}
\label{tab:translation-prompts}
\end{table}

\subsection{Experience and Results}

\begin{table}[htb]
    \centering
    \small
    \begin{tabular}{|l|l|l|}
    \hline
    \textbf{Error}                             & \textbf{Type}         & \textbf{Fixed} \\ \hline
    Missing BGP local-as attribute             & Syntax error          & Yes             \\ \hline
    Invalid syntax for prefix lists            & Syntax error          & Yes \\ \hline
    Missing/extra BGP route policy             & Structure mismatch   & Yes             \\ \hline
    Different OSPF link cost                   & Attribute error       & Yes             \\ \hline
    Different OSPF passive interface setting   & Attribute error       & Yes             \\ \hline
    Setting wrong BGP MED value                & Policy error & Yes             \\ \hline
    Different prefix lengths match in BGP      & Policy error & No \\ \hline
    Different redistribution into BGP          & Policy error & No              \\ \hline
    \end{tabular}
    \vspace{-1em}
    \caption{Translation errors found and whether GPT-4 was able to fix them with generated prompts.}
    \label{tab:translation-errors}
\end{table}\

We tried translating a Cisco configuration from the Batfish examples~\cite{batfish} into Juniper format. This configuration was short enough to fit within GPT-4 text input limits, but used non-trivial features including BGP, OSPF, prefix lists, and route maps. GPT-4's synthesized Juniper router configuration had several errors. In many cases, when an automatically generated prompt, similar to those in Table~\ref{tab:translation-prompts}, is provided to GPT-4, it will produce a response fixing the issue. In some cases, GPT-4 is unable to edit the translation correctly, either applying no change or applying an erroneous one. This often requires manual intervention via more specific prompts in order to fix. Another problem is that GPT-4 can fix one error, but introduce new errors that were not previously there. Sometimes it even reintroduces errors that were previously fixed! However, we were able to reach a point where Campion and Batfish no longer produced errors.

{\bf Leverage:} The entire cycle of prompts was around 2 human prompts and 20 automated prompts, for a leverage of 10X. Some of the 20 automatic prompt correction cycles included minor cycles for syntax correction not just at the start but also 
after correcting semantic errors.  To be clear, we ``simulated'' each API call by feeding our automatically generated prompts manually to GPT-4.

Table~\ref{tab:translation-errors} shows errors in the translation at some point and whether GPT-4 was able to fix them using an automatically generated prompt. In more detail:

\textbf{Missing BGP local-as attribute:} The translated BGP neighbor declarations  did not include a local AS attribute. We label this a syntax error since it produces a parse warning.

\textbf{Missing/extra BGP routing policy:} An import or export policy is used for a BGP neighbor in only one configuration. 

\textbf{Different OSPF link attributes:} OSPF links have a number of attributes, and the translation sometimes contains differences in link cost or passive interface settings.

\textbf{Setting wrong BGP MED value:} The translation of one BGP routing policy did not update the BGP MED value. This was caused by an error in translating one of the route map clauses from the original Cisco configuration.

\textbf{Different Redistribution behavior into BGP:} Cisco and Juniper formats handle route redistribution into BGP differently. Juniper typically does this using the same routing policies that control importing and exporting BGP routes while Cisco configurations set a separate route map for route redistribution. In our case, Campion detected that the Juniper configuration was redistributing some routes that the Cisco configuration did not. This could be fixed by adding a "from bgp" condition to a number of locations in the policy. Unlike the previously described errors, GPT-4 was unable to fix this when given the automatically generated prompt. Instead it usually does nothing when asked to fix the error. However, it was able to fix the problem when asked more directly to add "from bgp" conditions to routing policies.

\textbf{BGP prefix list issues:} Another subtle issue occurred when translating prefix lists. In the original Cisco configuration, a prefix list was defined to match prefixes with length 24 or greater where the first 24 bits matched 1.2.3.4. In Cisco this is done with the command:
\medskip

\noindent
\textit{\small ip prefix-list our-networks seq 5 permit 1.2.3.0/24 ge 24}

\medskip
\noindent
and it was applied with the definition:
\medskip

\noindent
\textit{\small route-map from\_customer deny 100} \\
\hspace*{1em} \textit{\small match ip address prefix-list private-ips}
\medskip

\noindent
The noteworthy part is the "ge 24" which says to match prefixes with length 24 or greater. There is no equivalent of this in defining prefix lists in Juniper, but for our use case, there are at least two methods of getting similar behavior in Juniper with different syntax. When GPT-4 is asked to translate the configuration, it often does not translate the "ge 24" part correctly, often just omitting it, so the space of prefixes matched will differ in the translation.  When asked to fix this problem, it sometimes generates configurations with incorrect syntax. For example, it can output the following:
\medskip

\noindent
\textit{\small
prefix-list our-networks \{ 1.2.3.0/24-32; \}
}
\medskip

\noindent
which is not valid Juniper syntax. However, after informing it of the error, it does eventually find a correct translation.

\section{Global Policies via Local Synthesis}
\label{sec:localsynthesis}
Next, we used GPT-4 to generate router configs for a given network topology based on local policies for each router, inspired by Lightyear~\cite{lightyear}, which does control plane verification by
verifying local invariants. We limited our scope to BGP.


For semantic correctness, we use two new modules. The first is a 'topology' verifier which checks whether the config of a particular router follows the defined topology. It checks whether GPT-4 sets up all  interfaces, declares BGP neighbors and announces networks correctly. Second, we run Batfish to check local policies defined in the prompts; the outputs are used to refine the result. 


\subsection{Method}

We begin by specifying the task to GPT in an initial prompt using a couple of sentences. The intention is to influence the LLM to start `thinking' in a certain fashion. Our goal is to make the network follow the no-transit policy, under which no two ISP's should be able to reach other. However, all ISPs should be able to reach the CUSTOMER and vice versa.
\begin{figure}[htb]
    \centering
    \includegraphics[width=0.8\linewidth]{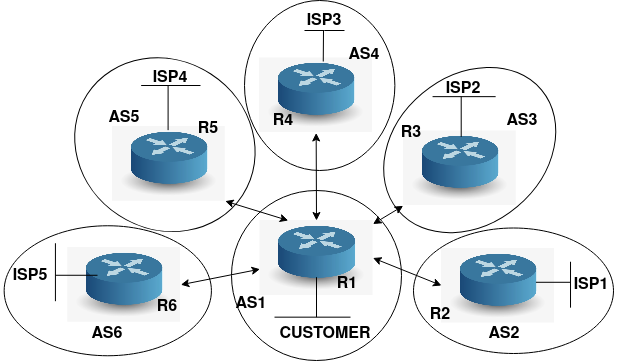}
    \vspace{-1em}
    \caption{Star network topology used for local synthesis.}
    \label{fig:topo}
\end{figure}


It is difficult to write a natural language description of the topology, a task prone to human error. 
We wrote an automated script that generates text given the topology as input. In our experiments, we limited our scope to star networks where one router would be attached to a CUSTOMER IP, while the other routers are connected to different ISPs (Figure~\ref{fig:topo}). All the ISP routers are directly connected to the first router. The "network generator" therefore only needs the number of routers as input. It has two outputs: 1) a textual description and 2) a JSON dictionary for the entire network topology. The textual description is used as a prompt, while the JSON dictionary is used later to check whether the generated configs match the topology.\\

{\bf Local versus Global Policy Prompts?}  We tried specifying to GPT-4 the global no-transit policy 
at once. GPT-4 generated two innovative strategies: filtering routes using AS path regular expressions, and  denying ISP prefixes from being advertised to other routers from the customer router. Unfortunately, we found after correcting topology and syntax errors, when we provided feedback in terms of a counterexample packet (as would be provided by a ``global" network verifier like Minesweeper), GPT-4 was confused and kept oscillating between incorrect strategies. We found that specifying local policies as in Lightyear~\cite{lightyear} gave us better results because it allowed us to localize verification errors to specific routers and specific route maps within those routers.  

We asked GPT-4 to generate configs for each router using a new prompt each time, specifying the {\em local policy} for each router.  Specifically, the policy is that $R1$ should add a specific community at the ingress to each ISP and then drop routes based on those communities 
at the egress to each ISP. The generated errors fell into three categories:

\textbf{Syntax errors: } GPT-4 generates a configuration with invalid Cisco syntax including errors in which certain config lines are misplaced. Batfish produces parse warnings identifying these errors.

\textbf{Topology errors: } GPT-4 incorrectly declares or misses some BGP neighbors or forgets to announce certain networks. For this, we use an automated "topology verifier" that compares the config against the previously specified JSON dictionary and outputs inconsistencies.

\textbf{Semantic errors / Policy errors: } GPT-4 produces configs that do not follow the intended local policy. We use Batfish "Search Route Policies" for verification in this step. In case there is a semantic error, Batfish produces an example where the local policy is not followed. This examples is then fed to GPT-4 in a fresh prompt.

Classifying into separate  categories allowed us to use different tools to address each one. Table ~\ref{tab:local_synth_errors} lists examples of the rectifying prompts. Once all the errors are rectified, we simulate the entire BGP communication using Batfish as a final step, in order to ensure that the global policy is satisfied, though the proof technique of Lightyear~\cite{lightyear} could instead be used to ensure that the local policies imply the global one.


\subsection{Experience and Results}



Since some GPT-4 errors were more common, we supplied it an IIP (the Inital Instruction Prompt) as follows:

{\em CLI prompts:} GPT-4 would often generate commands to enter on the Cisco command line interface, which is undesirable. Thus we specifically asked it to generate the .cfg files.

{\em Wrong keywords:} While generating the configs, it would often use certain keywords such as `exit', `end', `configure terminal', `ip routing', `write', `hostname' and `conf t'. It had a tendency to place some of them in the wrong locations. 
Hence, we directed it not to use these keywords. Any extra required commands for setup were prepended to the final config files, before running them on Batfish. 


{\em Match Community:} 
When trying to match against a community, it sometimes generates syntax like: \\
    {\em \small route-map \ \ FILTER$\_$ROUTES \ \ permit \ \ 10} \\
    \hspace*{1em} \textit{\small match community 100:1}

\noindent This is incorrect. The correct way to match against a community in a route-map is to first declare a community list that contains the community  as in:

   {\em \small ip community-list 1 permit 100:1}

\noindent and then while matching, make a call to a community list:

    {\em \small route-map FILTER\textunderscore ROUTES permit 10} \\
    \hspace*{1em}    {\em \small match community 1}

\noindent Thus we included another IIP to define a community list and then in a route-map, match using only this list. 
    
{\em Adding Communities:} While adding communities using a route-map, GPT-4 tends to generate syntax similar to: \\
    {\em \small route-map ADD\textunderscore COUMMUNITY permit 10} \\
    \hspace*{1em} {\em \small set community 100:1}

We observed that this happens even when we explicitly ask it to `add' a community to the route. The above route-map replaces all the communities that are already present in the route with the community 100:1. 
So we added an initial prompt saying that it should always use the 'additive' keyword when adding a community to the route.

These initial prompts along with the syntax rectification scheme of Table~\ref{tab:local_synth_errors} are able to eliminate common syntax errors produced by GPT-4. Despite this, we found two egregious cases where human intervention is needed:

\textbf{Placing neighbor commands in the wrong location:} In a config file for BGP, all network and neighbor commands must be placed under the "router bgp" block. For example, the neighbor command is used to attach a route-map to the ingress or egress of an interface. We found that in rare situations, GPT-4 defines a route-map and then associates it with an interface outside the "router bgp" block. Batfish is able to catch this syntax error. However, the output is not informative enough for GPT-4 to be able to fix the issue.


\textbf{AND/OR Semantics in match statements: } GPT-4 does not understand the semantic difference between placing multiple match conditions under a single route-map stanza versus placing them in different stanzas. For no-transit, we had asked GPT-4 to generate a config for $R1$ that would add a different community to every route incoming from $R2-R6$ (Figure~\ref{fig:topo}). We also asked it to filter routes containing any such community on the egress of the interface connecting $R1$ to $R2-R6$.  GPT-4 added the correct communities at the ingress, but
at the egress at $R1$ it incorrectly used AND semantics for filtering routes as in the following route-map for the $R1-R2$ interface:

{\em \small
\begin{lstlisting}[]
ip community-list 1 permit 100:1
ip community-list 2 permit 101:1
ip community-list 3 permit 102:1
ip community-list 4 permit 103:1
ip community-list 5 permit 104:1

route-map FILTER_COMM_OUT_R2 deny 10
 match community 2
 match community 3
 match community 4
 match community 5
route-map FILTER_COMM_OUT_R2 permit 20
\end{lstlisting}
}

Community 100:1 is associated with routes incoming from R2, 101:2 with those coming from R3 and so on. We desire routes incoming from $R3-R6$ to be filtered out at the egress to $R2$. The above config will only filter out routes that have all the communities 101:1, 102:1, 103:1 and 104:1, not any one of them. When we asked Batfish whether the above route-map filters routes that have the community 101:1, it produced a counterexample but this counterexample to GPT-4 failed to rectify the issue. A human prompt was needed to ask GPT-4 to declare each match statement in a separate route-map stanza. Our attempts to help GPT-4 distinguish between AND and OR semantics using an example in the IIP also failed.

\begin{table}[tp]
\small
\begin{tabular}{|l|l|l}
\cline{1-2}

\textbf{Type} &
  \textbf{Examples} &
   \\ \cline{1-2}
\begin{tabular}[c]{@{}l@{}}Syntax\\ error\end{tabular} &
  \begin{tabular}[c]{@{}l@{}}`\textit{ip community-list standard} \\ \textit{COMM\textunderscore LIST\textunderscore R2\textunderscore OUT permit .+}' is wrong syntax. \end{tabular} &
   \\ \cline{1-2}
\begin{tabular}[c]{@{}l@{}}Topology \\ error\end{tabular} &
  \begin{tabular}[c]{@{}l@{}}
    1. Interface \textit{eth0/1} ip address does not match with \\ given config. Expected \textit{2.0.0.1}, found \textit{2.0.0.2} \\
    2. Local AS number does not match. Expected \textit{1}, found \textit{3} \\
    3. Router ID does not match with given config. \\ Expected \textit{1.0.0.2}, found \textit{1.0.0.1} \\
    4. Neighbor with IP address \textit{1.0.0.1} and \\ AS \textit{1} not declared \\
    5. Network \textit{1.0.0.0/24} not declared \\
    6. Incorrect network declaration. \textit{7.0.0.0/24} is \\ not directly connected to \textit{R1} \\
    7. Incorrect neighbor declaration. No neighbor \\ with IP address \textit{7.0.0.2} AS \textit{7} found
  \end{tabular} &
   \\ \cline{1-2}
\begin{tabular}[c]{@{}l@{}}Semantic\\ error\end{tabular} &
  \begin{tabular}[c]{@{}l@{}}The route-map \textit{DROP\textunderscore COMMUNITY} permits routes \\ that have the community \textit{100:1}. However, they should \\ be denied. \end{tabular} &
   \\ \cline{1-2}
\end{tabular}
\vspace{-1em}
\caption{Sample rectification prompts for local synthesis. Batfish or the topology verifier fill in italicized text.}
    \label{tab:local_synth_errors}
\end{table}

{\bf Leverage:} The entire cycle took 2 human prompts and 12 automated prompts, for a leverage of 6X. 
Note that the AND-OR problem required a final correction prompt.

\section{Previous Work}
\label{sec:previouswork}
Jigsaw~\cite{jigsaw} and Copilot~\cite{copilot} use large language models for program synthesis.  While they concentrate on 
sequential programs, the deeper difference is that they do not pair the synthesizer with verifiers.  Jigsaw~\cite{jigsaw} does ask users to provide test cases and tests (but does not verify) the synthesized program.  Jigsaw also does some form of automatic syntax correction doing AST-to-AST transformations.  CoPilot~\cite{copilot} can suggest invariants
but does not attempt an axiomatic proof.  Jigsaw and Copilot do not address two questions we do: how to use
a specification, and how to provide localized feedback. 

The use of ChatGPT with the Kani Rust verifier~\cite{kani} comes closest to our vision; the Kani blog post finesses the  specification question (as we do for Cisco to Juniper) by focusing on {\em program transformations} (in their case optimization) for which the source program is the specification.  They also do not use modularity or local specifications.  More fundamentally the Kani~\cite{kani} use case does not do prompt programming: the user {\em always} manually switches between the verifier and the LLM, precluding possible leverage.

\section{Conclusions}
\label{sec:conclusions}
Our experiments are very preliminary but suggest:

{\em 1. Ramanujam Effect:} As with the mathematician Ramanujam, some of whose conjectures were incorrect and needed Hardy's help~\cite{hardy} for proofs, GPT-4 by itself is not ready for use without a verifier, making elementary errors that can
bring networks down.  

{\em 2. Verified Prompt Programming:} Using a verifier and automated corrections via a humanizer, 
GPT-4 can synthesize reasonable but not completely correct configurations for simple use cases, but the leverage in
reduced human effort can be high (5X to 10X). Modular verification seems crucial.




{\em 4. Local versus Global Specifications:} Modular synthesis is the dual to modular verification.  The search space for the LLM is large, which increases the chance that it will not be able to correctly complete a synthesis task based on a global specification. Instead the user needs to decide and describe the "roles" each node plays in satisfying the global spec.

 Much further testing in more complex use cases is needed.  Can GPT-4 add a new policy incrementally without interfering with existing verified policy?  While our paper is set in the context of network configuration, the vision, definitions (e.g., leverage) and lessons (e.g., the need for actionable local feedback, modularity, humanizers and IIPs) seem more generally useful to synthesize other programs.



\bibliographystyle{abbrv} 
\begin{small}
\bibliography{hotnets22}
\end{small}

\end{document}